\documentclass[10pt,aps,prb,twocolumn,superscriptaddress, nolongbibliography]{revtex4-2}
\usepackage{amsmath}

\usepackage{lmodern}
\usepackage[T1]{fontenc}
\setcitestyle{super}

\usepackage[utf8]{inputenc}
\usepackage{graphicx}
\usepackage{xcolor}

\usepackage[colorlinks=true,bookmarks=false,citecolor=blue,urlcolor=blue,linkcolor=red]{hyperref}

\begin{document}

\title{Passive harmonic mode-locked laser on lithium niobate integrated photonics}

\author{
Yu Wang$^{1,2,\ast}$, Guanyu Han$^{1,2,\ast}$, Jan-Philipp Koester$^{3}$, Hans Wenzel$^{3}$, Wei Wang$^{1,2}$, Wenjun Deng$^{1,2}$, Ziyao Feng$^{1,2}$, Meng Tian$^{1,2}$, Andrea Alù$^{1,2}$, Andrea Knigge$^{3}$, and Qiushi Guo${^{1,2,\dagger}}$\\
\textit{
$^1$Photonics Initiative, Advanced Science Research Center, City University of New York, NY, USA. \\
$^2$Physics Program, Graduate Center, City University of New York, 365 5th Ave, New York, 10016, NY, USA.\\
$^3$Ferdinand-Braun-Institut (FBH), Gustav-Kirchhoff-Str.4, 12489 Berlin, Germany.\\
$^\ast$These authors contributed equally to this work.}\\
$^\dagger$Email:
\href{mailto:qguo@gc.cuny.edu}{qguo@gc.cuny.edu}
}

\date{\today}

\begin{abstract}
Mode-locked lasers (MLLs) are essential for a wide range of photonic applications, such as frequency metrology, biological imaging, and high-bandwidth coherent communications. The growing demand for compact and scalable photonic systems is driving the development of MLLs on various integrated photonics material platforms. Along these lines, developing MLLs on the emerging thin-film lithium niobate (TFLN) platform holds the promise to greatly broaden the application space of MLLs by harnessing TFLN 's unique electro-optic (E-O) response and quadratic optical nonlinearity. Here, we demonstrate the first electrically pumped, self-starting passively MLL in lithium niobate integrated photonics based on its hybrid integration with a GaAs quantum-well gain medium and saturable absorber. Our demonstrated MLL generates 4.3-ps optical pulses centered around 1060 nm with on-chip peak power exceeding 44 mW. The pulse duration can be further compressed to 1.75 ps via linear dispersion compensation. Remarkably, passive mode-locking occurs exclusively at the second harmonic of the cavity free spectral range, exhibiting a high pulse repetition rate $\sim$20 GHz. We elucidate the temporal dynamics underlying this self-starting passive harmonic mode-locking behavior using a traveling-wave model. Our work offers new insights into the realization of compact, high-repetition-rate MLLs in the TFLN platform, with promising applications for monolithic ultrafast microwave waveform sampling and analog-to-digital conversion.

\end{abstract}

\maketitle

%\tableofcontents

\maketitle

%TC: ignore

%TC:endignore

\begin{figure*}[ht]
\centering
\includegraphics[width=0.92\linewidth]{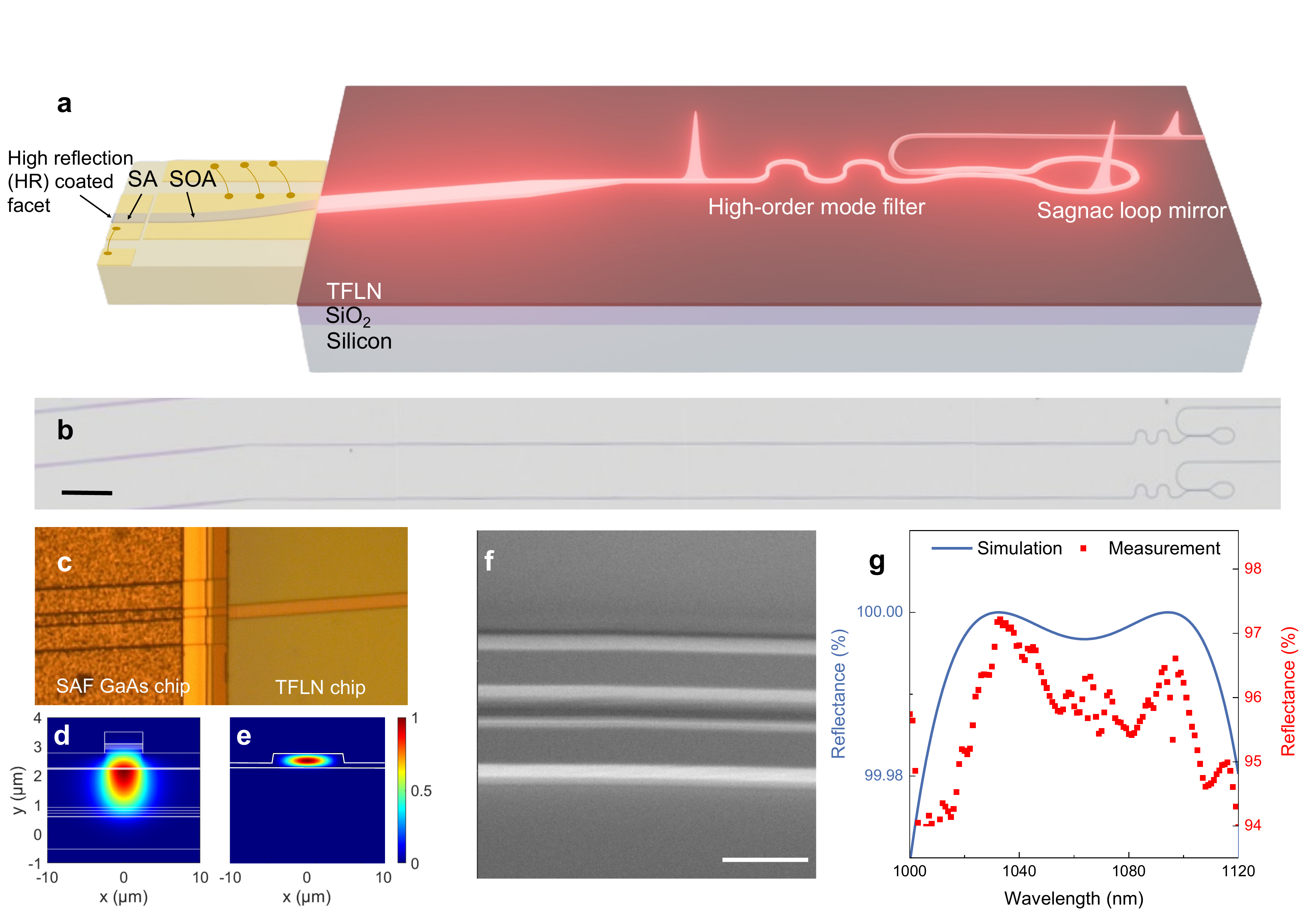}
\caption{\textbf{Hybrid GaAs–TFLN passive harmonic MLL} 
\textbf{a}, Schematic of the passive harmonic MLL. The GaAs RSOA chip contains an SOA section and an SA section, and it is butt-coupled to a TFLN extended cavity to form a Fabry-Perot laser cavity. When passively mode-locking occurs, two ultrashort light pulses exist within one cavity round-trip. \textbf{b}, Optical microscope image of the TFLN nanophotonic waveguide forming a passive extended cavity. Scale bar: 100 \textmu m. \textbf{c}, Butt-coupling interface between the GaAs RSOA chip and the TFLN chip. \textbf{d}, Fundamental TE mode in the GaAs waveguide at 1060 nm. \textbf{e} Fundamental TE mode in the input TFLN waveguide at 1060 nm.  \textbf{f}, SEM image of the CDC in the broadband Sagnac loop mirror. Scale bar: 2 \textmu m. The coupler has a total coupling length of 54 \textmu m and a gap of 600 nm between the waveguide top surfaces. \textbf{g}, Simulated (blue solid line) and measured (red symbols) reflection spectrum of the Sagnac loop mirror.}
\label{Fig1}
\end{figure*}

Mode-locked lasers (MLLs), with their distinctive features such as picosecond or femtosecond-scale output pulse durations, high peak intensities, high coherence, and broad spectral coverage, have enabled transformative advances across diverse fields. These include ultrafast laser
spectroscopy\cite{maiuri2019ultrafast}, supercontinuum generation {\cite{alfano1970emission,alfano1970observation,dudley2002supercontinuum}, optical atomic clocks\cite{bloom2014optical,nicholson2015systematic}, optical frequency combs and
frequency metrology\cite{picque2019frequency,cundiff2003colloquium}, biological imaging\cite{denk1990two,hell1996three,drexler2008optical}, to name a few. With the ongoing transition of photonic systems from bulky table-top setups to fully integrated platforms, realizing MLLs on integrated photonic circuits has become increasingly important, and has attracted growing interest\cite{liu2019high,srinivasan2015hybrid,dong20201,vissers2022hybrid,hermans2021high,winkler2024chip}. Among the various integrated photonic material platforms, developing integrated MLL in the thin-film lithium niobate (TFLN) platform is of particular significance. By synergistically combining MLLs with the TFLN's exceptional electro-optic (E-O) properties\cite{wang2018integrated,yu2022integrated,xu2020high,zhang2019broadband,hu2022high,feng2024integrated} and quadratic ($\chi^{(2)}$) optical nonlinearity\cite{wang2018ultrahigh,jankowski2020ultrabroadband,lu2019periodically,ledezma2022intense}, it may be possible to realize a suite of monolithic ultrafast photonic systems with performance and functionalities that are otherwise difficult to achieve in other platforms. For example, the co-integration of MLL with E-O devices on TFLN may enable fast wavelength tuning\cite{snigirev2023ultrafast, siddharth2025ultrafast}, active wavelength locking\cite{ma2024integrated}, and emerging applications such as high sampling rate, low-jitter microwave analog-to-digital converters (ADCs)\cite{kartner2008photonic}, ultra-low noise microwave sources\cite{kartner2019integrated}, and multi-wavelength coherent transmitters\cite{liu2019high,yang2022multi}. Moreover, on-chip all-optical switching\cite{guo2022femtojoule}, supercontinuum generation\cite{jankowski2023supercontinuum,hamrouni2024picojoule}, and high-rate quantum states generation\cite{nehra2022few,asavanant2019generation,javid2023chip} can be realized by directly interfacing the intense ultrashort pulses of an MLL with the strong $\chi^{(2)}$ nonlinear processes recently demonstrated in quasi-phase-matched TFLN nanophotonics. 

Despite recent advances in actively MLLs in TFLN~\cite{guo2023ultrafast}, the realization of high-repetition-rate, passively MLLs in TFLN remains elusive, yet warrants investigation for several reasons. First, an intra-cavity passive mode-locking element, such as a saturable absorber, allows for the generation of shorter pulses than its active counterparts due to the significantly shorter net gain windows\cite{ippen1994principles,kurtner2002mode}. Second, a passively MLL eliminates the need for external high-frequency microwave drivers, thereby significantly reducing system complexity. Passive mode-locking can also overcome the bandwidth limitation of a microwave driver, allowing for the generation of a high repetition rate pulse train. In this work, we present the first electrically pumped, self-starting passive harmonic MLL on TFLN. The MLL consistently and exclusively exhibits mode-locking at $\sim$20 GHz, which is the second harmonic of the cavity free spectral range (FSR). Importantly, such a harmonic passive mode-locking behavior enables a substantial increase in repetition rate without requiring cavity length reduction, which is highly favorable for developing high-repetition-rate on-chip MLLs. We also develop a traveling-wave model combined with a fast–slow carrier reservoir framework to elucidate the complex interplay between gain and saturable absorber dynamics, which underlies the observed self-starting passive harmonic mode-locking behavior.

\vspace{1.3mm}

\begin{figure*}[ht]
\centering
\includegraphics[width=0.9\linewidth]{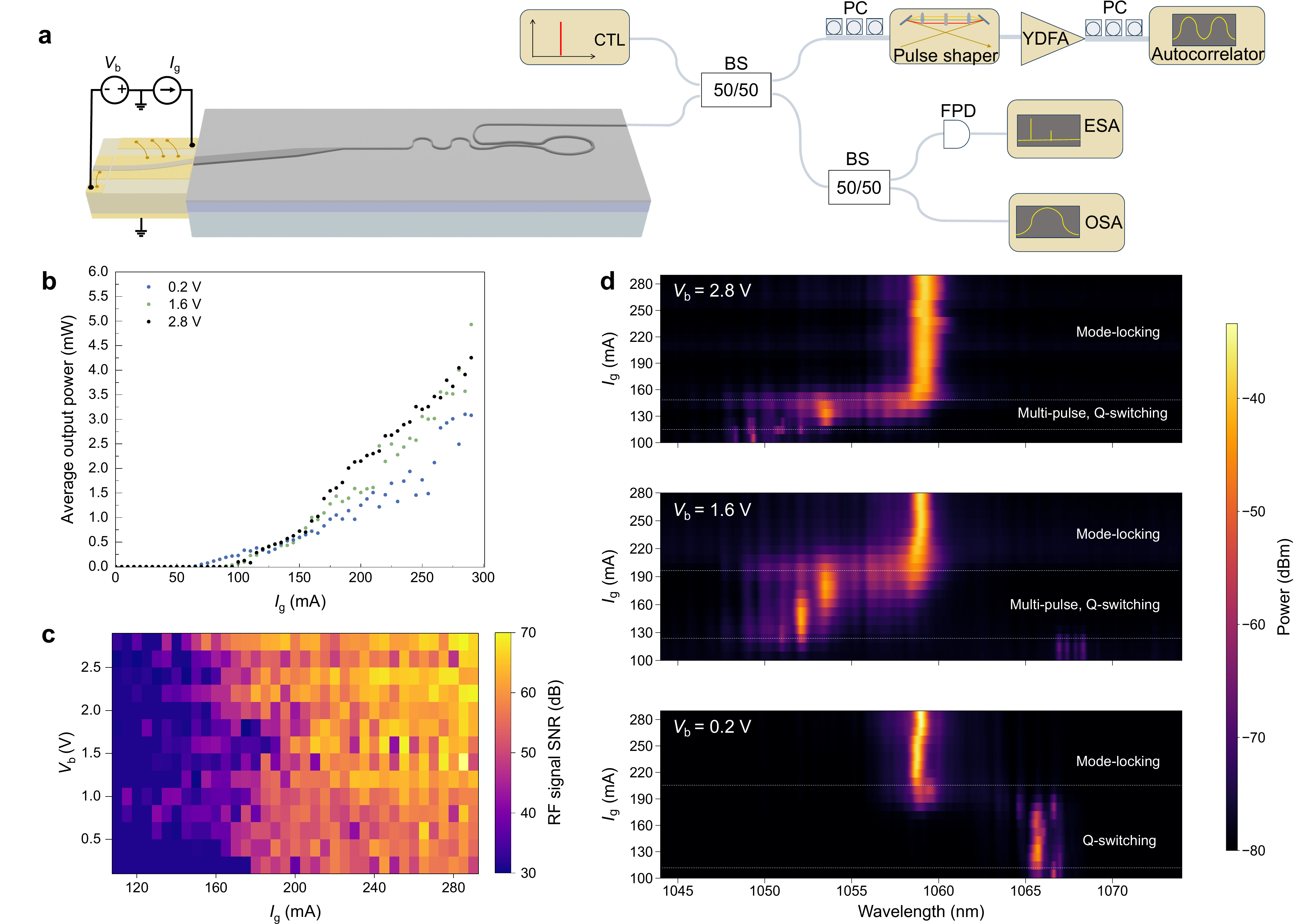}
\vspace{-5pt}
\caption{\textbf{Bias-dependent operating regimes of the passive harmonic MLL} \textbf{a}, Schematic of the MLL measurement setup. The laser signal emitted from the output facet was collected using a single-mode lensed fiber and divided by beam splitters (BS) for simultaneous monitoring of the output power with a power meter (PM), the optical spectrum with an optical spectrum analyzer (OSA), the RF spectrum with an electrical signal analyzer (ESA) through a fast photodetector (FPD), and the pulse intensity autocorrelator. Additionally, we used a continuous tunable CW laser (CTL) for heterodyne beat note measurement. Before the autocorrelator, we employed a pulse shaper for dispersion compensation and a ytterbium-doped fiber amplifier (YDFA). \textbf{b}, Output optical power as a function of gain current ($I_\mathrm{g}$) and bias voltage ($V_\mathrm{b}$). The lasing threshold currents are 65, 85, and 95 mA for $V_\mathrm{b} = 0.2$, 1.6, and 2.8 V, respectively.   \textbf{c}, SNR of RF signal around 20 GHz under various $I_\mathrm{g}$ and $V_\mathrm{b}$.  \textbf{d}, Laser spectra under various $V_\mathrm{b}$ and $I_\mathrm{g}$. White dotted lines separate different operating regimes.
}\label{Fig2}
\end{figure*}

Figure~\ref{Fig1}a shows the schematic of the passive harmonic MLL, which was realized through the hybrid integration of a single-angled-facet (SAF) GaAs reflective semiconductor optical amplifiers (RSOA) chip operating around 1060 nm and an especially designed TFLN extended cavity. The GaAs RSOA chip comprises a $\sim$1.8-mm-long electrically pumped semiconductor optical amplifier (SOA) section and a 185-\textmu m-long saturable absorber (SA) section upon reverse bias voltage for passive mode-locking. As shown in the optical microscopic image in Fig. \ref{Fig1}b, the fabricated TFLN extended cavity comprises a 0.7-mm-long angled tapered waveguide, a 216-\textmu m-long adiabatic taper, a 2-mm-long waveguide with 700 nm top width, a higher-order mode filter, and a broadband Sagnac loop mirror. A Fabry-Perot laser cavity configuration is formed between the reflective rear facet of the GaAs chip and the Sagnac loop mirror on the TFLN chip. Details about the GaAs chip and the TFLN extended cavity fabrication can be found in the Methods.

 The coupling regions between the GaAs RSOA and the TFLN chips are shown in Fig. \ref{Fig1}c. To minimize coupling loss, the input facet of the TFLN extended cavity was tapered to 9 \textmu m in top width and angled at 5$^\circ$. Such a design ensures a maximal modal overlap between the 1060 nm fundamental TE-modes in the GaAs waveguide (Fig.~\ref{Fig1}d) and that of the TFLN waveguide (Fig.~\ref{Fig1}e) at the corresponding facets, thereby achieving the highest optical transmission. This design yields a calculated single-pass chip-to-chip coupling loss of 3.55 dB, accounting for modal mismatch and the Fresnel reflection at the interface (see Supplementary Information Section 1 for details). Following the 9-\textmu m-wide input waveguide, we designed a 216-\textmu m-long adiabatic taper that gradually reduces the waveguide's top width from 9 \textmu m to 700 nm, with a negligible loss of 0.03 dB according to simulations. To ensure stable single-spatial-mode lasing, the 700-nm-wide waveguide also incorporates a higher-order mode filter consisting of four consecutive 20-\textmu m-radius semi-circular bends, which introduce an additional single-pass bending loss of 13.05 dB to the first-order TE mode compared to the fundamental.

At the end of the TFLN extended cavity, we implemented a Sagnac loop mirror with a curved directional coupler (CDC)\cite{mitarai2020design,morino2014reduction} to achieve a broadband reflection and high fabrication error tolerance. Figure~\ref{Fig1}f shows the scanning electron microscope (SEM) image of the CDC region, which features a 600 nm gap and 600 \textmu m bending radius, and a total coupling length of 54 \textmu m. As shown in Fig.~\ref{Fig1}e, the measured reflection spectrum of the Sagnac loop mirror (red symbols) closely resembles the simulated reflection spectrum (blue curve). Notably, due to the CDC, the Sagnac loop mirror not only features the broadband reflection centered around 1060 nm, but also exhibits slightly reduced reflection at the peak gain wavelength ($\sim$1060 nm) and enhanced reflection at nearby wavelengths around 1020 nm and 1090 nm. This reflection profile, in turn, results in a broadening of the round-trip gain spectrum and facilitates shorter pulse generation. Detailed discussion of the Sagnac loop mirror design methodology, structural parameters, and evaluation results are provided in the Supplementary Information Section 2. For the entire TFLN extended cavity, we estimate a round-trip group delay dispersion (GDD) of approximately 368.75 fs$^2$. Considering the entire laser cavity with both the GaAs and TFLN sections, the total cavity FSR is estimated to be around 10.02 GHz. Calculation details are provided in Supplementary Information Section 3.

\begin{figure*}[ht]
\centering
\includegraphics[width=0.95\linewidth]{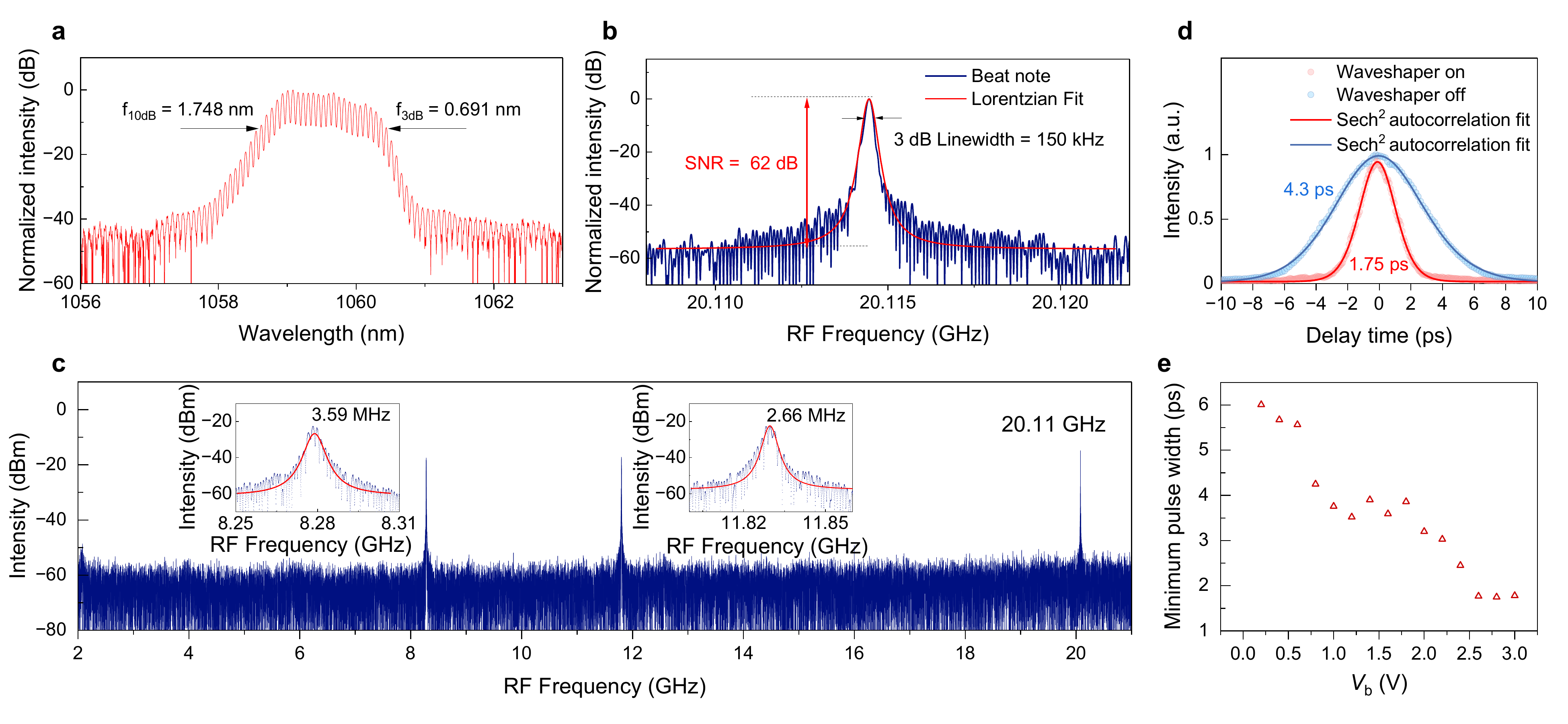}
\vspace{-10pt}
\caption{\textbf{Characterization of second harmonic passively MLL} \textbf{a}, Laser output spectrum measured at $V_\mathrm{b} = $2.8 V and $I_\mathrm{g}$ = 290 mA. The 10 dB bandwidth is approximately 1.748 nm, containing 23 comb lines due to axial modes. The 3 dB bandwidth is around 0.691 nm by fitting with the Fourier transform of a Sech$^2$ function. \textbf{b}, Normalized RF signal (blue) and its Lorentzian fit (red) centered around 20.115 GHz, with a fitted 3 dB linewidth of 150 kHz and a SNR of 62 dB. \textbf{c}, Heterodyne RF beat note measurement showing three spectrally clean and sharp peaks at 8.28, 11.83, 20.11 GHz. Insets: Zoomed-in view of the heterodyne beat notes. Blue symbols are measured data, and red lines are Lorentzian fits. \textbf{d}, Autocorrelation traces together with Sech$^2$ autocorrelation fits of the pulse at $V_\mathrm{b} = 2.8$ V. The corresponding pulse durations are 1.75 ps with the waveshaper on (red) and 4.3 ps with the waveshaper off (blue), obtained by multiplying the fitted FWHM values of the autocorrelation traces by the deconvolution factor 0.6482. \textbf{e}, Minimum pulse widths after dispersion compensation at different $V_\mathrm{b}$ from 0.2 V to 2.8 V. } 

\label{Fig3}
\end{figure*}

We characterized the mode-locking behavior using the experimental setup shown in Fig.~\ref{Fig2}a, which simultaneously collects the laser output power, the optical spectra, the RF spectra, and time-domain autocorrelation traces. The SOA section of the GaAs chip was driven by an injection current $I_\mathrm{g}$, while the SA section was reverse-biased at a voltage $V_\mathrm{b}$. During the measurement, we optimized the alignment between the two chips by maximizing the photocurrent measured at the SA section. The maximum photocurrent of $\sim$34 mA was obtained at $I_\mathrm{g}$ = 290 mA. Figure \ref{Fig2}b shows the measured on-chip laser output average power as a function of $I_\mathrm{g}$ and $V_\mathrm{b}$. As $V_\mathrm{b}$ increases from 0.2 V to 2.8 V, the lasing threshold current rises from 65 to 95 mA, indicating increased absorption of the SA with higher reverse voltage bias. Moreover, the laser exhibited the highest on-chip average lasing power of 5.03 mW at $I_\mathrm{g}$ = 290 mA and $V_\mathrm{b}$ = 1.6 V.
\vspace{1.3mm}

In parallel, we recorded the RF and optical spectra of the MLL to identify the mode-locking states. Interestingly, under all bias conditions, no RF signal around 10 GHz was observed, meaning that the fundamental mode-locking state is absent. Instead, we only found a strong peak near 20 GHz, suggesting the presence of a second harmonic mode-locking regime. Figure~\ref{Fig2}c presents the signal-to-noise ratio (SNR) of 20 GHz RF signals under various bias conditions. The SNR generally increases with $I_\mathrm{g}$ due to higher lasing power; however, at higher $V_\mathrm{b}$, the laser can enter the mode-locking regime at even lower $I_\mathrm{g}$. This observation indicates that a higher reverse bias of the SA can facilitate the passive mode-locking. Further analysis of the RF spectra shows that an SNR above $\sim$55 dB is typically associated with a clean single peak near 20 GHz, indicative of stable passive harmonic mode-locking. Moreover, when $I_\mathrm{g}$ decreases to the boundary of the mode-locking state, in some circumstances, two sidebands emerge around the central RF peak around 20 GHz, indicating the onset of Q-switching behavior, where a slow envelope modulates the pulse train in the time domain. In the Q-switched mode-locking regime, the device exhibits reduced stability and occasionally switches to multi-pulse operation (see Supplementary Information Section 4 for details).
\vspace{1.3mm}

Figure \ref{Fig2}d shows the evolution of the output optical spectra of the MLL as a function of 
$V_\mathrm{b}$ and $I_\mathrm{g}$, in which transitions between multi-mode continuous-wave (CW) lasing, Q-switched mode-locking, and mode-locking regimes can be seen. Consistent with RF spectrum SNR analysis, the $I_\mathrm{g}$ threshold for entering the stable mode-locking regime reduces as $V_\mathrm{b}$ increases from 0.2 to 2.8 V. Moreover, the optical spectral bandwidth of the MLL significantly broadens with increasing $V_\mathrm{b}$, suggesting a trend toward generating shorter pulses. This behavior is further corroborated by the increasing number of comb lines within the 10 dB bandwidth—from 5  to 23—as $V_\mathrm{b}$ increases from 0.2 to 2.8 V. We attribute the observed increased lasing threshold, enhanced RF signal SNR, and broader optical spectra at a higher $V_\mathrm{b}$ to the quantum confined Stark effect of semiconductor SAs\cite{prziwarka2017generation,miller1984band}: At higher $V_\mathrm{b}$, the absorption and extinction ratio increase, and the carrier recovery time decreases\cite{karin1994ultrafast}, thus leading to more stable mode-locking and shorter pulse generation. At lower $I_\mathrm{g}$, Q-switched mode-locking and multi-mode CW lasing regimes emerge, with central wavelengths shifting either toward shorter or longer wavelengths. 

Next, we examine the performance of the MLL at a representative operating point (\( I_\mathrm{g} = 290 \)~mA, \( V_\mathrm{b} = 2.8 \)~V), where the laser operates stably in the passive harmonic mode-locking regime for several hours. As shown in Fig.~\ref{Fig3}a, the laser output exhibits a flat-top, comb-like spectrum peaked at 1059 nm, with a 10 dB bandwidth of 1.748 nm. The comb line spacing is approximately 75.4 pm, corresponding to a repetition rate of 20.11 GHz. We fitted the optical spectrum with the Fourier transform of a Sech$^2$ function, as it is the solution ansatz of passively MLL with a slow saturable absorber\cite{haus2003theory}. The fitting yields a 3 dB bandwidth of 0.691 nm (Supplementary Information 4). The corresponding RF spectrum (Fig.~\ref{Fig3}b) further confirms the purely second harmonic passive mode-locking behavior: we only observed a strong single RF peak at 20.11 GHz with an SNR of 62 dB and 3 dB linewidth of 150 kHz, while the 10 GHz fundamental peak remained absent. We then quantify the coherence of the resulting MLL comb lines by collecting the heterodyne RF beat notes between two neighboring laser emission lines and a narrow linewidth ($\sim$10 kHz) reference CW tunable laser (CTL1050, Toptica). As shown in Fig. \ref{Fig3}c, two spectrally narrow RF heterodyne beat notes at $f_\mathrm{1}$ = 8.28 GHz and $f_\mathrm{2}$ = 11.83 GHz were obtained, with a full-width at half-maximum (FWHM) linewidth of 3.59 MHz and 2.66 MHz, respectively.

Complementing the spectral characterization, we characterized the temporal characteristics of the MLL's output pulses using an intensity autocorrelator. As shown in Fig.~\ref{Fig3}d (blue symbols and solid line), the autocorrelation trace reveals a pulse width of 4.3 ps, corresponding to a deduced on-chip pulse width of 4.29 ps, accounting for the estimated $4.84 \times 10^5~\mathrm{fs}^2$ GDD introduced by $\sim$20 m long optical fiber in the measurement setup (See Supplementary Information Section 5 for detailed calculation). Considering the 0.691 nm 3 dB bandwidth of the optical spectrum, the time-bandwidth product is calculated to be 0.79, which is 2.51 times larger than the 0.315 time-bandwidth product for a Sech$^2$ pulse. This suggests the presence of a large chirp imposed on the on-chip pulses. To determine the minimum achievable pulse width of our laser, we inserted a pulse shaper before the autocorrelator to compensate for dispersion. As shown in Fig. \ref{Fig3}d (red curve), the minimum pulse width of 1.75 ps was achieved when a large linear dispersion compensation of around 5 ps/nm was applied, corresponding to a GDD of $-2.65 \times 10^6\ \mathrm{fs}^2$ at a center wavelength of 1060 nm. 

Given the small round-trip GDD ($\sim$338 fs\textsuperscript{2}) of the TFLN extended cavity, we attribute the large linear chirp on the pulses primarily to the combined effects of strong self-phase modulation (SPM) and group-velocity dispersion (GVD) introduced by the SOA and SA sections \cite{gee1997intracavity}. When a short pulse enters the SOA section, its leading edge encounters undepleted carriers, while its trailing edge encounters depleted carriers. The carrier depletion increases the refractive index of the SOA via the linewidth enhancement factor \cite{henry1982theory,henry1981spectral}. This leads to a strong SPM and a quadratic frequency chirp across the pulse: the leading edge becomes red-shifted, while the trailing edge is blue-shifted\cite{delfyett1994femtosecond}. Simultaneously, the pulse acquires a large linear chirp (temporal broadening), as the leading edge propagates faster than the trailing edge \cite{agrawal2002effect}. Upon entering the SA section, the nonlinear chirp of the pulse is largely canceled. Consequently, the pulse exhibits a dominant large linear chirp, with a value of $\sim$1.8 $\times$ 10$^6$ fs$^2$}/mm as reported in Ref.\cite{agrawal2002effect}, depending on wavelengths and laser structures. Experimentally, we found the linear chirp shows no strong dependence on $V_\mathrm{b}$ and $I_\mathrm{g}$, and the applied dispersion compensation for achieving the shortest pulse width is always around 5 ps/nm. Moreover, applying nonlinear dispersion compensation did not further shorten the pulse. Figure~\ref {Fig3}e shows the minimum achievable pulse width decreases as $V_\mathrm{b}$ increases, which is consistent with the spectral broadening effect shown in Fig.~\ref {Fig2}d. Moreover, we observed no strong correlation between pulse width and gain current.
\vspace{1.0mm}

\begin{figure*}[ht]
\centering
\includegraphics[width=0.95\linewidth]{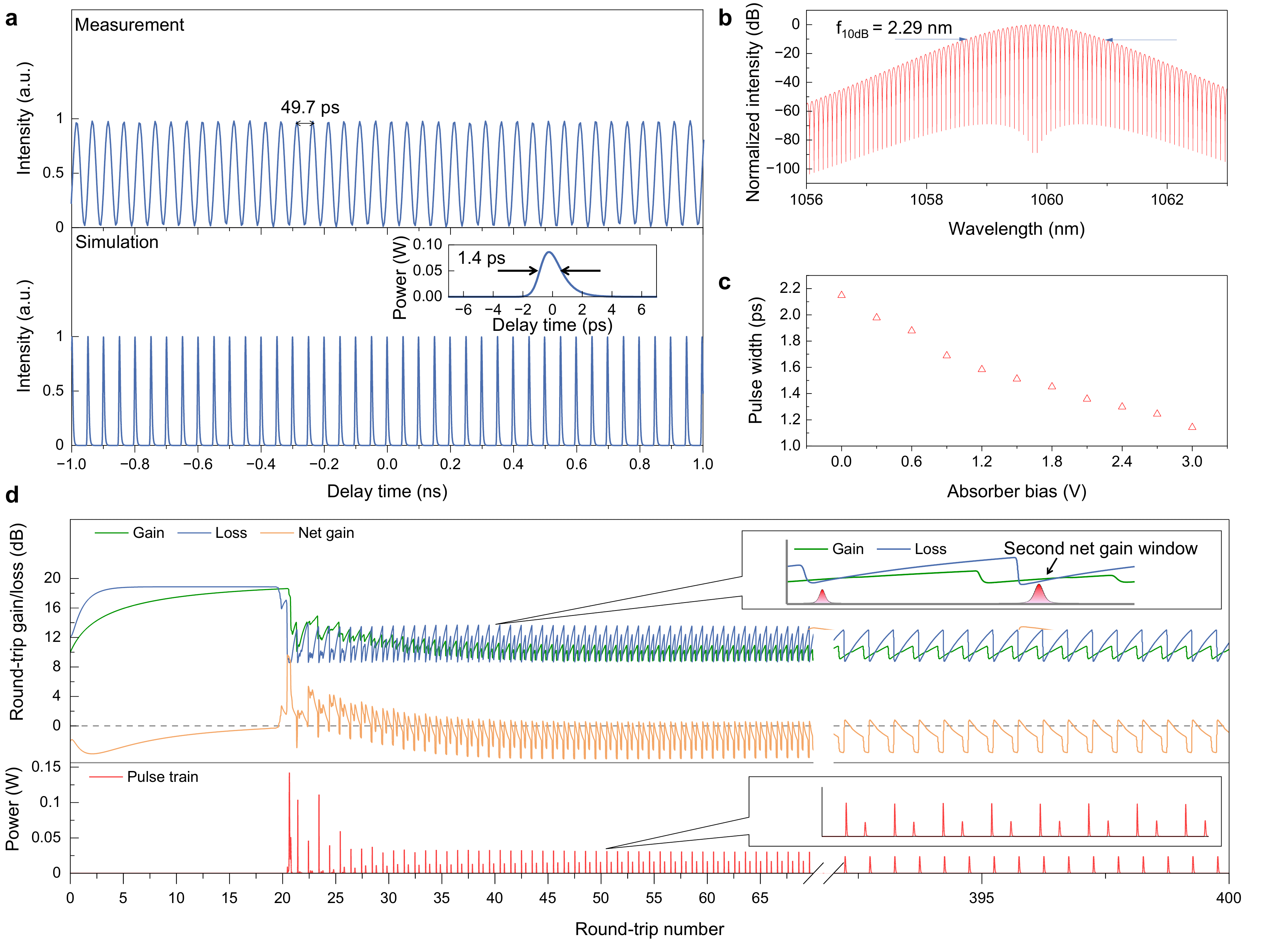}
\vspace{-5pt}
\caption{\textbf{Passive harmonic MLL dynamics} \textbf{a}, Comparison of experimental and simulated pulse trains. Top: measured pulse train using a fast photodetector and a high-speed oscilloscope; Bottom: simulated pulse train. Inset: Zoom-in of simulated pulse exhibiting a steep leading edge and a prolonged trailing tail at $V_\mathrm{b}$ = 2.8 V, $I_\mathrm{g}$ = 260 mA.
\textbf{b}, Simulated laser optical spectrum showing comb line spacing of approximately 75 pm and a 10 dB bandwidth of 2.29 nm. \textbf{c}, Simulated pulse width as a function of $V_\mathrm{b}$ at $I_\mathrm{g}$ = 290 mA. \textbf{d}, Upper: Evolution of round-trip gain, loss, and net gain. Inset: Zoom-in view of the gain and loss dynamics responsible for the formation of the second net gain window. Lower: Round-trip pulse train formation. Inset: Zoom-in view of early round-trips showing a pulse and its satellite, compared with a uniformly spaced pulse train at steady state.
}\label{Fig4}
\end{figure*}

As shown in Fig. \ref{Fig4}a, the temporal pulse train captured by the fast photodetector and the high-speed oscilloscope directly confirms the second harmonic passive mode-locking at 20.1 GHz repetition rate. To gain deeper insight into the harmonic mode-locking dynamics, we employed a traveling wave model that separates the laser into two distinct regions: an active region comprised of SOA, SA, and the gap sections; and a passive TFLN extended cavity, which only imposes linear dispersion and linear loss. Detailed parameters of our model are shown in the Supplementary Information Section 6. The temporal and spatial evolutions of the optical fields in both SOA and SA are governed by the partial differential equations\cite{cuyvers2021hybrid}:

{
\small

\begin{equation}
\frac{\partial A^{\pm}(z,t)}{\partial t} \pm \frac{\partial A^{\pm}(z,t)}{\partial z} =
j \frac{\omega_\mathrm{0} v_\mathrm{g} \Gamma}{2 n_{\text{eff}} c} \chi(\omega, N_\mathrm{f}) A^{\pm}(z,t)
- \beta \, A^{\pm}(z,t),
\end{equation}
}

\noindent where \( A^{+}(z,t) \) and \( A^-(z,t) \) represent the forward and backward propagating optical fields, and \( \chi(\omega, N_{\mathrm{f}}) \) denotes the complex susceptibility, which depends on both the angular frequency \( \omega \) and the carrier density in the active region \( N_\mathrm{f} \).  \( \omega_\mathrm{0} \), \( n_\mathrm{eff} \), \( v_\mathrm{g} \), \(\Gamma\), and \( \beta \) are the central angular frequency, effective refractive index, group velocity, optical confinement factor, and linear loss, respectively. We model the carrier dynamics within the GaAs quantum well using a two-reservoir framework, where we divide carriers into a fast active region and a slow reservoir\cite{sun2005gain,eisenstein1989gain}:

\begin{align}
\frac{\partial{N_\mathrm{f}}}{\partial{t}} &= \frac{\eta_\mathrm{i} I_\mathrm{f}}{qV_\mathrm{f}} + \frac{\Gamma \, \mathrm{Im}[\chi(\omega, N_\mathrm{f})]}{n_{\text{eff}} c} |A|^2 + \frac{N_\mathrm{s} - N_\mathrm{f}}{\tau_{\text{t}}(1+\eta_\mathrm{p})} \\
\frac{\partial{N_\mathrm{s}}}{\partial{t}} &= \frac{\eta_\mathrm{i} I_\mathrm{s}}{qV_{\mathrm{s}}} - \frac{N_\mathrm{s}}{\tau_\mathrm{life}} - \frac{N_\mathrm{s} - N_\mathrm{f}}{\tau_{\text{t}}(1+\eta_\mathrm{p})},
\end{align}

\noindent where \( N_\mathrm{s} \) is the density of the slow carriers (in the unit of $\mathrm{cm^{-3}}$) near the active region, associated with the current injection and diffusion processes occurring on the nanosecond timescale (\( \tau_\mathrm{life} \)). \( N_\mathrm{f} \) refers to the density of fast carriers in the active region, which participate directly in the stimulated emission process. The total current applied to the gain \( I_\mathrm{g} \) is split into \( \eta_\mathrm{i} I_\mathrm{f} \) and \( \eta_\mathrm{i} I_\mathrm{s} \), corresponding to the currents injected into the fast and slow reservoirs, respectively. \( \eta_\mathrm{i} \) is the current injection efficiency\cite{coldren2012diode}. \( V_\mathrm{f} \) and \( V_\mathrm{s} \) are the active volumes of the fast and slow regions, respectively. The transition time between the two reservoirs is \( \tau_{\mathrm{t}} \), which is typically in the order of picoseconds\cite{alfieri2017optical}, and \( \eta_\mathrm{p} \) is the population inversion factor. A smaller value of \( \eta_\mathrm{p} \) indicates a faster gain recovery, as it reflects a larger reservoir of carriers available to replenish the active region. \( q \) represents the charge of an electron.

Notably, as shown in Fig. \ref{Fig4}a lower panel, the simulated pulse trains using this model closely reproduce the second harmonic mode-locking state observed experimentally at an $I_\mathrm{g}$ = 290 mA and $V_\mathrm{b}$ = 2.8 V. The Fourier transform of the pulse train (Fig.~\ref{Fig4}b) reveals a spectrum with a comb line spacing of 75 pm, corresponding to a repetition rate of approximately 20 GHz. Moreover, as shown in Fig. \ref{Fig4}c, the model captures the observed reduction in pulse width with increasing SA reverse-bias voltage $V_\mathrm{b}$, arising from two combined effects: (1) the absorber recovery time decreases exponentially with $V_\mathrm{b}$, and (2) the initial absorption increases approximately linearly with $V_\mathrm{b}$\cite{kim1996reverse}. These effects jointly narrow the net gain window for pulse formation, enabling the generation of shorter pulses.

The self-starting harmonic passive mode-locking behavior originates from the large transient net gain window created by the fast gain recovery and resulting in unsaturated gain of the GaAs quantum wells, which enables the formation of multiple optical pulses within a single cavity round-trip. To elucidate this mechanism, we simulated the evolution of the round-trip gain and loss by monitoring their dynamics at the mid-point of the SOA and SA, together with the pulse train at the output. The detailed calculation of the gain and loss is provided in Supplementary Information Section 8. As shown in Fig. \ref{Fig4}d, the large reservoir of unsaturated carriers at start-up amplifies random background perturbations, leading to the formation of an irregular, chaotic waveform. After a few round-trips, the saturable absorber efficiently suppresses weaker pulses, allowing only the dominant pulse and, in certain conditions, its satellite to survive and stabilize in the cavity. Later, the interplay between gain and loss reaches equilibrium, forming a stable two-net-gain-window regime. The satellite pulse escapes further from the main pulse due to its higher group velocity, causing the two pulses to evolve toward equal temporal spacing and amplitude gradually. This ``self-adjusting'' process is driven by differential carrier depletion between pulses: the stronger pulse induces greater carrier depletion, locally increasing the group index and slowing down the mode. In comparison, the weaker pulse causes less depletion and advances relative to the stronger pulse. After many more round-trips, this dynamic ultimately leads to a stable, evenly spaced second harmonic mode-locking state, and the two pulses become identical in intensity. We verified this behavior by extracting the SOA carrier-induced changes in group indices and performing a similar analysis of the fundamental mode-locking regime, as detailed in Supplementary Information Section 8.

In summary, we demonstrate an electrically pumped, passive harmonic MLL on a lithium niobate photonic integrated circuit, which generates 4.3 ps pulses with a peak power of 44.4 mW around 1060 nm on chip. The output pulse width can be further compressed to 1.75 ps with external dispersion compensation. Our MLL exhibits a stable second-harmonic mode-locking state at a 20 GHz repetition rate, effectively overcoming the repetition rate limitation imposed by the physical cavity length. Such a self-starting, passive second harmonic mode-locking behavior arises from the unsaturated gain in the GaAs SOA and a subsequent self-adjusting process of the intra-cavity pulses, which are well captured by our traveling-wave model. We envision that shorter output pulses, higher peak power, and improved coherence of the MLL can be achieved by integrating other components available on TFLN, such as $\chi^{(2)}$ saturable absorbers\cite{guo2022femtojoule} or intra-cavity E-O modulators for hybrid mode-locking\cite{delfyett1994femtosecond}. Moreover, the large gain dispersion arising from the GaAs SOA can be compensated by chirped multi-waveguide gratings\cite{huang2024thin,liu2024ultra,zhang2020integrated} while maintaining a compact device footprint. To achieve even higher-order harmonics and higher repetition rates, two or more saturable absorbers can be integrated at different positions within the laser cavity, enabling regimes such as colliding pulses\cite{marsh2017mode,li2010harmonic}, which have been shown to boost the repetition rate of MLL beyond 100 GHz. By seamlessly integrating our MLL with electro-optic modulators on TFLN, its high repetition rate can be harnessed to realize monolithic RF ADCs with ultra-high sampling rates and low timing jitter\cite{kartner2008photonic}.

%TC:ignore
\section*{Methods}
\noindent \textbf{Device fabrication.} 
The GaAs-based two-section RSOA used in this work was fabricated at the Ferdinand-Braun-Institut. The vertical waveguide is realised as an asymmetric large optical cavity (ASLOC) design comprising Al$_{0.35}$Ga$_{0.65}$ and Al$_{0.85}$Ga$_{0.15}$As n- and p-cladding layers surrounding a 2.4 \textmu m-thick Al$_{0.25}$Ga$_{0.75}$As waveguide core that embeds a compressively strained InGaAs single quantum well having GaAsP spacer layers. The resulting refractive index profiles of the core and cladding layers are shown in Supplementary Information Section 1.  The device consists of a bent 1785 \textmu m-long SOA section and a straight 185 \textmu m-long SA section located at the rear facet. These sections are electrically insulated by a 30 \textmu m-wide insulation gap in the p-metallization and by shallow He$^{+}$ implantation, which enables a forward bias current at the SOA while the SA operates under a reverse voltage bias. A highly reflective (95 \%) coating was applied to the rear facet to form one end of the Fabry–Perot-type extended laser cavity. The front facet of the RSOA features an anti-reflection coating and is angled at 3° to suppress parasitic back reflections. 
\vspace{1.3mm}

The fabrication of the TFLN extended cavity was accomplished at the CUNY ASRC nanofabrication facility. We fabricated the TFLN waveguides on a 600-nm-thick X-cut magnesium-oxide (MgO) doped TFLN on a SiO$_2$/Silicon (4.7 \textmu m/500 \textmu m) substrate. The waveguide patterns were defined using a 100 keV e-beam lithography system (Elionix ELS-G100) with hydrogen silsesquioxane (HSQ) as the e-beam resist. Ar+ ion-beam Inductively Coupled Plasma (ICP) etching was then used to etch 400 nm of thin film, forming a 60° sidewall angle. The remaining resist was stripped using buffered oxide etch (BOE). Both edges of the chip were mechanically polished to improve coupling efficiency at the input and output waveguide facets.
\vspace{1.3mm}

\noindent \textbf{Optical measurements.} The reflectance spectrum of the Sagnac loop mirror shown in Fig. 1g was measured using a supercontinuum source (NKT Photonics, SuperK FIANIUM). Light was coupled into the input facets of the LN chip via a single-mode lensed fiber (OZ Optics) and the transmitted light was collected by another single-mode lensed fiber. Input and output powers were monitored with power meters (Thorlabs PM20). With the alignment optimized and fixed, we used a tunable filter (NKT Photonics, SuperK SELECT) to tune the laser wavelength from 1000 to 1120 nm in 1 nm steps. For the characterization of the MLL, the GaAs chip was mounted on a 3-axis stage (Thorlabs NanoMax), with its position finely adjusted using a three-channel piezo controller (Thorlabs MDT693B). The SOA section was driven by a current source (Thorlabs LDC220C), and the SA section was reverse-biased using a voltage source meter (Keithley 2602b). The laser output was collected via a single-mode lensed fiber (OZ Optics), with the output power monitored on a power meter (Thorlabs PM20) before being directed to a fast photodetector (Thorlabs RXM25AF). The detector output was fed into an electrical signal analyzer (Keysight N9030A) and a high-speed oscilloscope (Keysight Infiniium UXR0254B) to record RF signals and pulse trains (Figs.~2c, 3b, and 3c). Optical spectra (Figs.~2d and 3a) were measured using an optical spectrum analyzer (YOKOGAWA AQ6370D). For intensity autocorrelation measurements, a Ytterbium-doped fiber amplifier (Thorlabs YDFA-SM) was used to amplify the collected laser signals. The autocorrelation trace shown in Fig.~3d was measured with an intensity autocorrelator (Femtochrome FR-103XL). A pulse shaper (Coherent Waveshaper 1000A/SP) was employed to apply linear dispersion compensation.
\vspace{1.3mm}

\noindent \textbf{Numerical simulations.} We used commercial software (Ansys Lumerical Inc.) to solve for the waveguide modes, calculate the dispersion properties, and design the broadband Sagnac loop mirror. In the simulation, the anisotropic refractive index of the LN was modeled by the Sellmeier equations\cite{zelmon1997infrared}. For the second harmonic mode-locking simulation, we used MATLAB to solve the traveling wave equations. We integrated the carrier rate equation for both SA and SOA into the electrical susceptibility. We used loss, linear, and quadratic phase shifts to represent the optical feedback from the passive waveguide. The details regarding the simulation method can be found in the Supplementary Information Section 6.

\section*{Data Availability}
The data that support the plots within this paper and other findings of this study
are available from the corresponding author upon reasonable request.
\section*{Code Availability}
The computer code used to perform the traveling wave simulations in this paper is available from the corresponding author upon reasonable request.
\section*{Acknowledgements}
The authors acknowledge support from NSF Grant No. 2338798 (CAREER Award), the Department of Defense, the start-up grants from the CUNY Advanced Science Research Center and the CUNY Graduate Center, and the PSC CUNY Research Award. Y.W. acknowledges support from the Graduate Fellowship and the CUNY Science Scholarship from the CUNY Graduate Center. The authors also thank Prof. Gabriele Grosso and Dr. John Woods for providing access to laboratory facilities.

\section*{Authors Contributions}
Y.W. and G.H. fabricated the TFLN chip and performed the measurements with assistance from W.W., W.D., and Z.F. J.-P.K. prepared and pre-characterized the GaAs RSOA chips. H.W. designed the GaAs chips. A.K. supervised the GaAs chip fabrication; G.H. developed the LN waveguide fabrication process. Y.W. developed the traveling-wave model, analyzed the experimental results, and performed the simulations with assistance from W.D. and M.T. Y.W. and Q.G. wrote the manuscript with input from all authors. Q.G. conceived and supervised the project. 
\section*{Competing Interests}
The authors declare no competing interests.

\newpage
\bibliographystyle{apsrev4-2}

\bibliography{references}

\end{document}